\documentclass[twoside,twocolumn,10pt]{article}
\usepackage[pdftex]{color}
\usepackage{wscg}
\usepackage{url}
\usepackage[english]{babel}
\usepackage{mathtools}
\usepackage{amsmath}
\usepackage{tabu,booktabs}
\usepackage[pdftex]{graphicx}
\graphicspath{{figures/}}
\usepackage{caption}
\DeclareCaptionType{copyrightbox}
\usepackage{subcaption}
\usepackage[numbers]{natbib}
\usepackage{microtype}

\ifdefined\anonymous
	\usepackage[switch]{lineno}
	\linenumbers
\else
	\usepackage{nopageno}
	\usepackage{flushend}
\fi

\usepackage[pdfusetitle]{hyperref}

\makeatletter
\def\affiliation#1{\def\@affiliation{#1}}
\def\@maketitle{%
  \begin{center}%
  \let \footnote \thanks
    \sffamily
    {\fontsize{16pt}{19.2pt} \bfseries \@title \par}%
    \vskip 1.0em%
    {%
      \lineskip .5em%
      \begin{tabular}[t]{c}%
        \@author
      \end{tabular}%
      \vskip 0.5em%
      \@affiliation%
      \par}%
  \end{center}%
  \par
  \vskip 0.5em}
\makeatother

\title{Evaluation of 4D Light Field Compression Methods}
\ifdefined\anonymous
	\author{Author 1 \and Author 2 \and Author 3 \and Author 4 \and Author 5}
	\affiliation{Faculty of ...\\University of ...\\Address ...\\Country ...\\\{author1,author2,author3,author5\}@domain}
\else
	\author{David Barina \and Tomas Chlubna \and Marek Solony \and Drahomir Dlabaja \and Pavel Zemcik}
	\affiliation{Centre of Excellence IT4Innovations\\Faculty of Information Technology\\Brno University of Technology\\Bozetechova 1/2, Brno\\Czech Republic\\\{ibarina,ichlubna,isolony,zemcik\}@fit.vutbr.cz}
\fi

\makeatletter
\def\Uslash{\mathbin{\mathchar`\/}\@ifnextchar{/}{\kern-.15em}{}}
\g@addto@macro\UrlSpecials{\do \/ {\Uslash}}
\def\Ucolon{\mathbin{\mathchar`:}\@ifnextchar{/}{\kern-.1em}{}}
\g@addto@macro\UrlSpecials{\do : {\Ucolon}}
\makeatother

\begin{document}

\twocolumn[{\csname @twocolumnfalse\endcsname

\maketitle

\begin{abstract}
Light field data records the amount of light at multiple points in space, captured e.g. by an array of cameras or by a light-field camera that uses microlenses.
Since the storage and transmission requirements for such data are tremendous, compression techniques for light fields are gaining momentum in recent years.
Although plenty of efficient compression formats do exist for still and moving images, only a little research on the impact of these methods on light field imagery is performed.
In this paper, we evaluate the impact of state-of-the-art image and video compression methods on quality of images rendered from light field data.
The methods include recent video compression standards, especially AV1 and XVC finalised in 2018.
To fully exploit the potential of common image compression methods on four-dimensional light field imagery, we have extended these methods into three and four dimensions.
In this paper, we show that the four-dimensional light field data can be compressed much more than independent still images while maintaining the same visual quality of a perceived picture.
We gradually compare the compression performance of all image and video compression methods, and eventually answer the question, "What is the best compression method for light field data?".
\end{abstract}

\subsection*{Keywords}
Light field, Plenoptic imaging, Lossy compression, Image refocusing

\vspace*{1.0\baselineskip}
}]

\section{Introduction}
\label{sec:Introduction}

\copyrightspace

\begin{figure}[t]
	\def\svgwidth{\linewidth}
	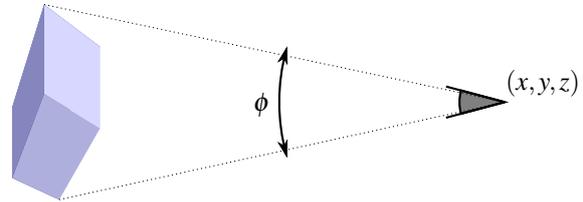
	\caption{
		Plenoptic capture of a scene from a single viewing position.
		For simplicity, the range of viewing angles is indicated for one spherical coordinate.
	}
	\label{fig:pleno1}
\end{figure}

To describe a three-dimensional scene from any possible viewing position at any viewing angle, one could define a plenoptic function
$P(x,y,z,\phi,\psi)$, where the $(x,y,z)$ is the position and $(\phi,\psi)$ is a viewing angle (in spherical coordinates) of a camera.
Figure~\ref{fig:pleno1} shows the situation.
The value of the $P$ is color.
The definition can be further extended with $t$ (time) to describe a dynamic scene.

\begin{figure}[t]
	\def\svgwidth{\linewidth}
	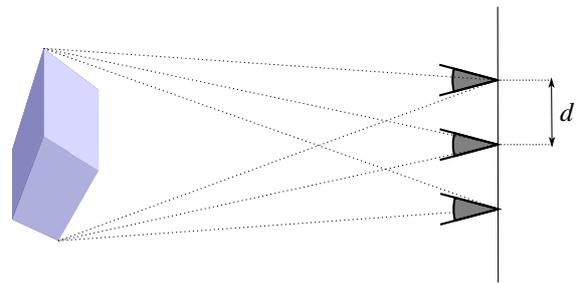
	\caption{4D light field capture via an array of cameras.}
	\label{fig:pleno2}
\end{figure}

\begin{figure}
	\centering
	\includegraphics[width=.5\linewidth]{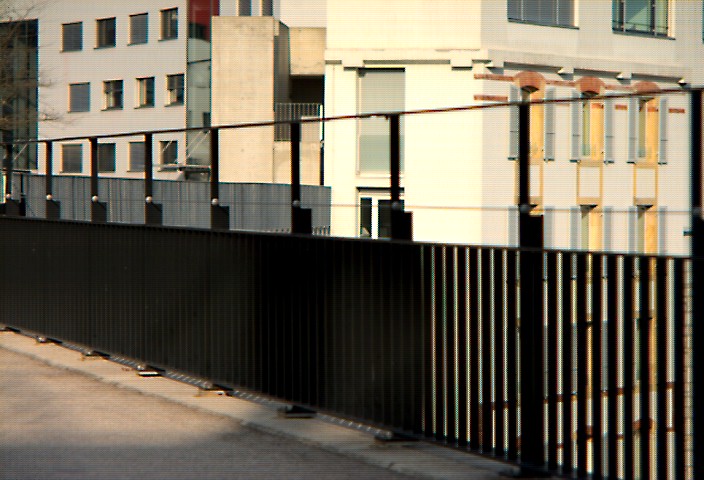}%
	\includegraphics[width=.5\linewidth]{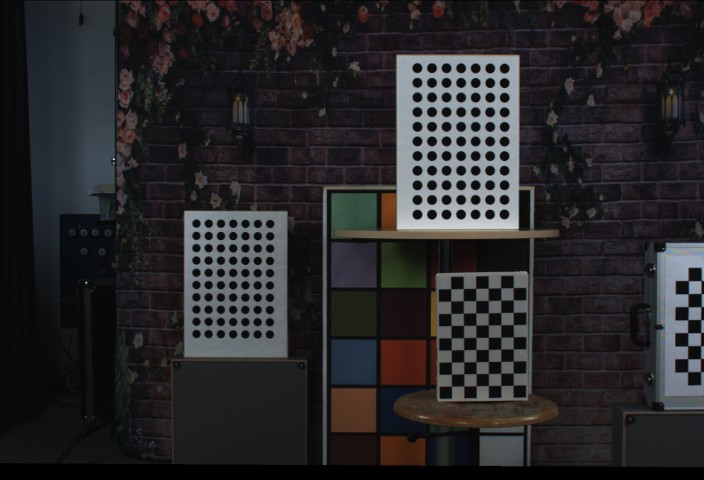}\\[-1pt]
	\includegraphics[width=.5\linewidth]{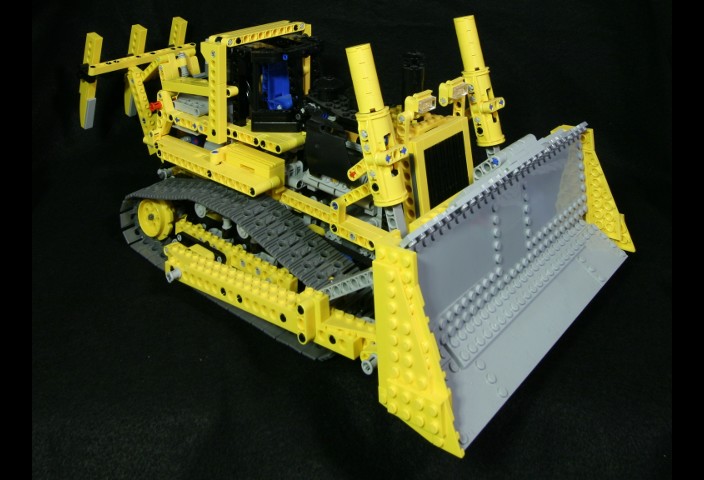}%
	\includegraphics[width=.5\linewidth]{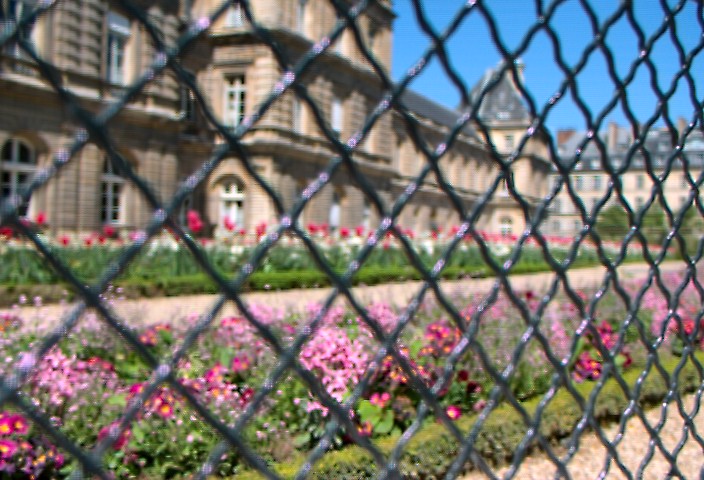}
	\caption{
		Dataset used in this paper.
		From top left to bottom right: Black Fence, Chessboard, Lego Bulldozer, and Palais du Luxembourg.
	}
	\label{fig:dataset}
\end{figure}

Our interest here is to describe the scene by capturing either via an array of cameras or by a single compact sensor preceded by microlenses.
In this case, the aperture is modeled by a grid of views (cameras) located on a two-dimensional plane.
This situation is shown in Figure~\ref{fig:pleno2}, where the baseline between individual views from the grid is described by the distance $d$.
This representation is often referred to as 4D light field (LF)
	since we deal with the light field function, $L$, sampled in four dimensions, $(k,l,m,n)$, where the $(m,n)$ are pixel coordinates, and $(k,l)$ are indices of a sub-aperture image.
Light fields acquired by the single compact sensor have limited support for the viewing angle.
Light fields based on the array of cameras offer larger viewing angles at the cost of missing information in between the cameras.
In practice, the number of views located on the two-dimensional plane ranges from a couple of units to several hundred.
Considering increasing resolution sensors, it is no surprise that the light field data reach huge sizes.
As an example, consider "Lego Bulldozer" light field (Figure~\ref{fig:dataset}) taken from the Stanford Light Field Archive.
The light field is captured using a $17\times17$ grid of cameras having image resolution $1536\times1152$ (rectified and cropped).
The uncompressed size easily exceeds a gigabyte.
For light field videos, storage and transmission requirements are enormous.

Several methods to compress 4D light fields have been recently proposed.
Some of them attempt to compress directly the data from sensors preceded by microlenses (lenslet image).
Other compresses the resulting 4D light field.
In this paper, we focus only on the latter ones.
We compare various state-of-the-art compression methods applicable to 4D light field data.
These methods include recent video compression standards, especially AV1 (validated in June 2018), and XVC (version released in July 2018).
In order to evaluate the comparison, we refocus the original and decompressed light field.
The evaluation is then carried out using the PSNR as a full-reference quality assessment metric.

The remainder of the paper is organized as follows.
Section \ref{sec:Related Work} reviews related work and compression methods.
Section \ref{sec:Evaluation} presents our experiments in detail, and discusses the results.
Section \ref{sec:Conclusions} concludes the paper.

\section{Related Work}
\label{sec:Related Work}

\begin{figure*}
	\def\svgwidth{\linewidth}
	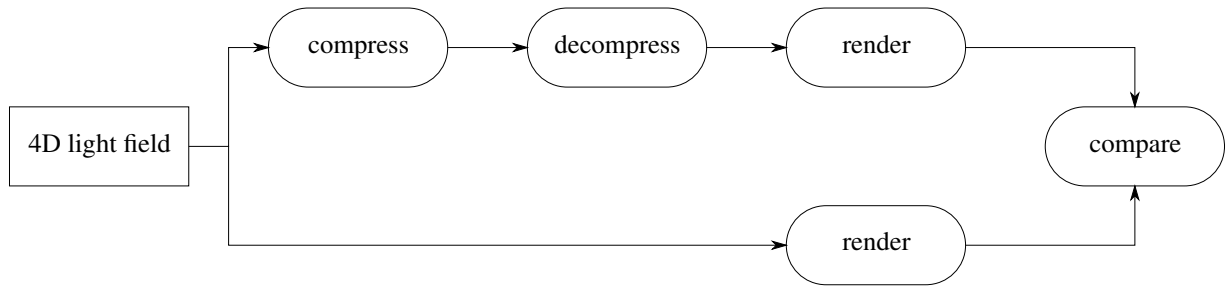
	\caption{Data flow diagram of the compression performance assessment methodology used in this paper.}
	\label{fig:dfd}
\end{figure*}

The individual views from a light field are usually never displayed.
Therefore, it is not very meaningful to compare the original and decompressed light field directly, even though such methodology is usual to asses a single view compression performance.
For this reason, we adopt the compression performance assessment methodology for multi-focus rendering from \cite{Alves2016}.
This methodology basically lies in assessing the quality of the rendered views for multiple focal points.
The rendered views are obtained by combining pixels from different 4D light field views for various focal planes.
The average distortion is computed as the mean of the PSNR for multiple rendered focal plane views.
This situation is shown in Figure~\ref{fig:dfd}.
Note that the PSNR is computed from the MSE over all three color components.

The 4D light field comprises a two-dimensional grid of two-dimensional views.
The baseline between individual views ranges from a few millimeters (microlenses) to several centimeters (camera array).
It is, therefore, natural to expect a high similarity of views adjacent in any of two grid directions.
This similarity opens the door to understanding the 4D light field data as a video sequence navigating between the viewpoints.
Another possible point of view is to see the 4D light field as the three- or directly four-dimensional body.
The above approaches can also be reflected in light field compression by using either an image, video, volumetric, or four-dimensional coding system.
Although other approaches (like 3D video) are also possible, we are not aware of generally available coding systems for such cases.

In recent years, several papers compared and evaluated the compression performance of various approaches on light field imagery.
The authors of \cite{Alves2016} evaluated the performance of the main image coding standards, JPEG, JPEG 2000, H.264/AVC intra profile, and H.265/HEVC intra profile.
The "intra" suffix refers to the fact the individual views were compressed independently (intra profile).
The video coding approaches were not evaluated.
As could be expected, the H.265/HEVC intra profile proved to be the most efficient compression method.
In \cite{Viola2017}, the authors compared the compression performance of three strategies using the H.265/HEVC.
Their first strategy performs compression directly on the lenslet image.
Another strategy arranges 4D LF views a pseudo-temporal sequence in spiral order and subsequently compressed it.
The last strategy compresses a subset of lenslet images through the transformation to 4D LF.
Their results show that coding 4D LF leads to better performance when compared to coding lenslet images directly.
The authors of \cite{Higa2013} compared the performance of JPEG, JPEG 2000, and SPIHT directly on lenslet images.
The comparison was performed using the same methodology as in this paper.
As could be expected, the JPEG 2000 exhibits the best compression performance.
In \cite{Perra2017}, the authors proposed to rearrange 4D LF views into tiles of a big rectangular image.
This image is then compressed using the JPEG 2000 coder.
The proposed scheme was compared against standard image coding algorithms, namely the JPEG 2000 and JPEG XR.
It is, however, unclear how these standard coding algorithms were exactly applied to the 4D light field data.
In \cite{Aggoun2006}, the author rearranges the 4D light field into a three-dimensional body.
The three-dimensional volume is then encoded using the 3D DCT scheme on $8\times8\times8$ blocks, similarly as in the JPEG coding system.

Besides conventional coding methods, also an alternative approach \cite{Bakir2018} exists that uses deep learning to estimate the 2D view from the sparse sets of 4D views.
Another approach \cite{Chen2018} proposes own sparse coding scheme for the entire 4D LF based on several optimized key views.
The method in \cite{Jiang2017} decomposes the 4D light field into homography parameters and residual matrix.
The matrix is then factored as the product of a matrix containing $k$ basis vectors and a smaller matrix of coefficients.
The basis vectors are then encoded using the H.265/HEVC intra profile.
In \cite{Li2017,Li2017b}, the authors propose a hierarchical coding structure for 4D light fields.
The 4D LF is decomposed into multiple views and then organized them into a coding structure according to the spatial coordinates.
All of the views are encoded hierarchically.
The scheme is implemented in the reference H.265/HEVC software.
In \cite{Han2018}, the authors propose a coding scheme that splits the 4D light field into several central views and remaining adjacent views.
The adjacent views are subtracted from the central views, and both groups are then encoded using H.265/HEVC coder.
The authors of \cite{Li2014,Li2016} feed the 4D LF into the H.265/HEVC exploiting the inter prediction mode for individual LF views.
Finally, tremendous attentions have also been focused on convolutional neural network based compression approaches \cite{Hou2019,Jia2018}.

From the above, it can be seen that the JPEG 2000 and especially the H.265/HEVC coding schemes are quite popular.
In this paper, we compare the compression performance of the main state-of-the-art lossy compression methods.
These methods can be divided into four groups according to the way they process the 4D LF data.
The first group covers the following image coding methods---the JPEG and JPEG 2000. 
In the literature \cite{Li2018}, this kind of methods is sometimes referred to as the self-similarity based methods.
The second group comprises video coding methods: H.265/HEVC, AV1, VP9, and XVC. 
In the literature, these methods are referred to as the pseudo sequence based methods.
The third group extends the image coding methods into three dimensions.
This group consists of JPEG 3D (our own implementation) and JPEG 2000 3D (Part 10, JP3D).
Notice that the JPEG 3D refers to a volume image rather than a pair of stereoscopic images.
The fourth group extends the image coding methods into four dimensions.
However, only one method in this group exists, JPEG 4D (our own implementation).
To evaluate the above methods, we use the following list of encoders: OpenJPEG, x265, libaom (AV1 Codec Library), libvpx (VP8/VP9 Codec SDK), xvc codec, and our own implementation of the JPEG method. 

Since our comparison also deals with the latest video compression standards, we consider it appropriate to present their short description here.
The H.265/HEVC (High Efficiency Video Coding, MPEG-H Part 2) is a video compression standard designed as a successor to the widely used H.264/AVC (MPEG-4 Part 10).
The standard was published in June 2013.
The AV1 (AOMedia Video 1) is an open video coding format standardized in June 2018.
It succeeds the VP9 video coding format developed by Google.
According to \cite{Zabrovskiy2018}, the AV1 outperforms the H.265/HEVC by 17\,\%, and VP9 by 13\,\% over a wide range of bitrate/resolutions.
The XVC is a video coding format with a strong focus on low bitrate streaming applications.
The official website claims that the codec outperforms the AV1, H.265/HEVC, and VP9.

\section{Evaluation}
\label{sec:Evaluation}

\begin{table*}
	\begin{tabu} to \linewidth {l|X[l]|c|l}
		\toprule
		description          & source                                & resolution                             & disparity \\
		\midrule
		Black Fence          & EPFL Light-field data set             & \makebox[3.4em]{$15 \times 15$}$\times$\makebox[5.3em]{$625 \times 434$}   & \makebox[1em]{$-1$} to $1$ \\
		Chessboard           & Saarland University                   & \makebox[3.4em]{$8 \times 8$}$\times$\makebox[5.3em]{$1920 \times 1080$}   & \makebox[1em]{$40$} to $90$ \\
		Lego Bulldozer       & Stanford Computer Graphics Laboratory & \makebox[3.4em]{$17 \times 17$}$\times$\makebox[5.3em]{$1536 \times 1152$} & \makebox[1em]{$-1$} to $7$ \\
		Palais du Luxembourg & EPFL Light-field data set             & \makebox[3.4em]{$15 \times 15$}$\times$\makebox[5.3em]{$625 \times 434$}   & \makebox[1em]{$-1$} to $1$ \\
		\bottomrule
	\end{tabu}
	\caption{
		Dataset used in this paper.
		The first and last light field are taken using a plenoptic camera; the Chessboard is captured using a camera array; the Lego Bulldozer is captured using a motorized gantry holding a camera.
		The adjacent image disparity range (last column) is given in pixels.
	}
	\label{tab:dataset}
\end{table*}

This section presents our dataset, multi-focus rendering method, experiments conducted on this dataset using the above methodology, and the results we achieved.

Our dataset consists of four 4D light fields based on two types of capturing devices.
Two of the light fields were captured using Lytro Illum B01 plenoptic camera and the other two using conventional cameras.
The first conventional camera light field was captured using a multi-camera array, and the other one using simple motorized gantry equipped with Canon Digital Rebel XTi camera.
Corresponding resolutions and adjacent image disparity ranges are listed in Table~\ref{tab:dataset}.
The value in the last column describes the pixel difference in the location of the same 3D object projected to images captured by a camera or computed from a lenslet image in the case of Lytro.
The range is narrow (ca. $-1$ to $+1$ pixel) for the densely-sampled light field (Lytro) and wide (ca. $40$ to $90$ pixels) for the images captured by camera array.
These values correlate to the focal length and camera baseline (distance between camera centers).
For convenience, the central view for each light field is shown in Figure~\ref{fig:dataset}.

The digital refocus of the images at the virtual focal plane is achieved using shift-sum algorithm \cite{Ng2005}.
This algorithm shifts the sub-aperture images (views) according to camera baseline with respect to the reference frame and accumulates the corresponding pixel values.
The refocused image will be an average of the transformed images.
The computation of the pixel value at point $(m, n)$ of the refocused image $E_d$ is given by the equation
\begin{align}
	E_d(m, n) = \frac{1}{N} \sum_{k, l} L(k, l, m + \alpha k, n + \alpha l),
\end{align}
where $N$ is the number of summed images,
$\alpha$ is the distance of the synthetic plane from the main lens,
$k$ and $l$ are indices of a sub-aperture image of the light field representation, and $\alpha k$ and $\alpha l$ are the shift parameters with respect to the reference frame.
We performed a linear interpolation in the last two 4D dimensions to convert the sampled light field function into a continuous one.

\begin{figure}
	\begin{subfigure}[b]{\linewidth}
		\includegraphics[width=\linewidth]{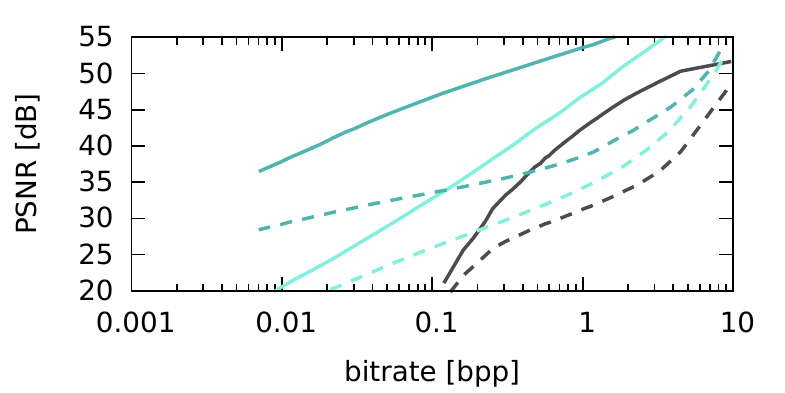}%
	\end{subfigure}%
	\\%
	\includegraphics[width=\linewidth,trim={0 0 0 1.75cm},clip]{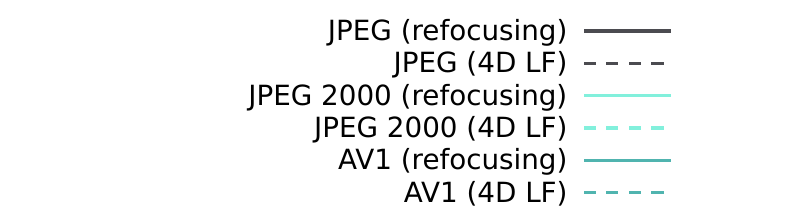}%
	\medskip%
	\caption{
		Experiment 0.
		The difference in the quality assessment using the 4D light field directly vs. using images rendered at virtual focal planes.
		Illustratively shown on the Black Fence light field.
	}
	\label{fig:experiment0}
\end{figure}

\paragraph{Experiment 0}
Before we start, the reader might wonder whether it is really necessary to assess the image quality on views rendered for multiple focal points rather than the original views (i.e. compare the original and decompressed LF directly).
A quick experiment reveals that a big difference exists between the former and the latter (see Figure~\ref{fig:experiment0}).
This difference is about 10 decibels in the PSNR, depending on the bitrate and compression method.
This can be explained by the fact that any pixel in the rendered view is a sum of pixels from the 4D LF so that this sum all together suppresses compression artifacts.
In other words, we can afford to compress the 4D light fields much more than independent images, while maintaining the same visual quality of a screened picture.

\begin{figure*}
	\begin{subfigure}[b]{.5\linewidth}
		\includegraphics[width=\linewidth]{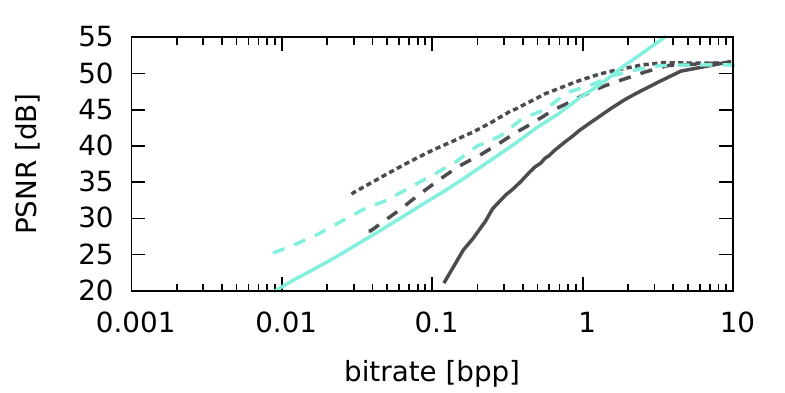}%
		\caption{Black Fence}
	\end{subfigure}%
	\begin{subfigure}[b]{.5\linewidth}
		\includegraphics[width=\linewidth]{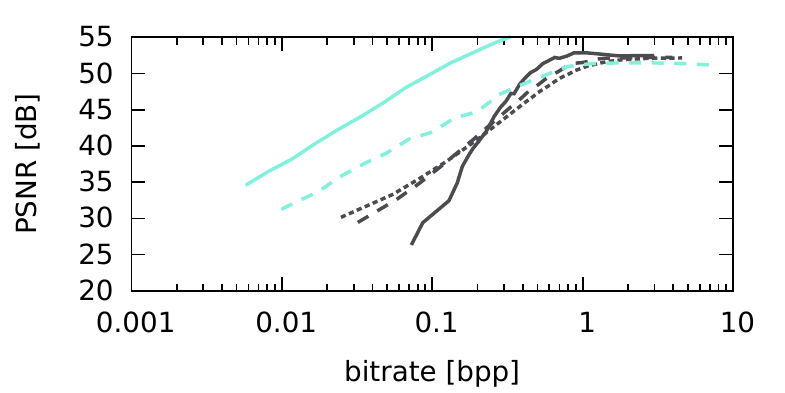}%
		\caption{Chessboard}
	\end{subfigure}%
	\\[10pt]%
	\begin{subfigure}[b]{.5\linewidth}
		\includegraphics[width=\linewidth]{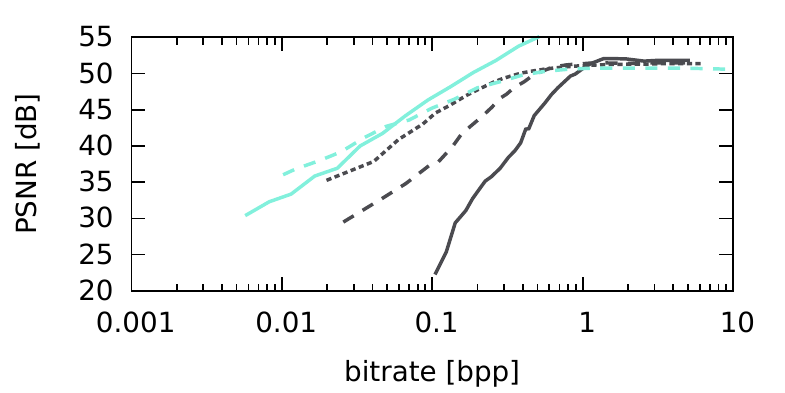}%
		\caption{Lego Bulldozer}
	\end{subfigure}%
	\begin{subfigure}[b]{.5\linewidth}
		\includegraphics[width=\linewidth]{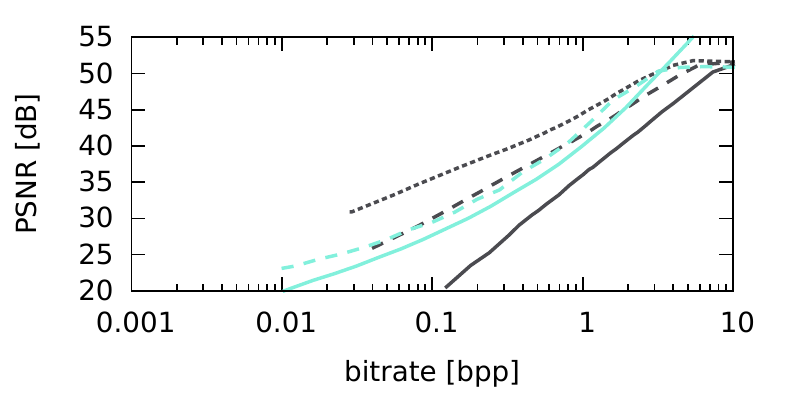}%
		\caption{Palais du Luxembourg}
	\end{subfigure}%
	\\[15pt]%
	\centering
	\includegraphics[width=.5\linewidth,trim={0 0 0 2.75cm},clip]{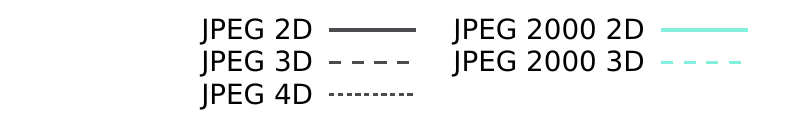}%
	\medskip%
	\caption{
		Experiment 1.
		Comparison of image compression methods against their extensions into three and four dimensions.
	}
	\label{fig:experiment1}
\end{figure*}

\paragraph{Experiment 1}
As seen from the previous section, most current LF compression approaches handle either 2D data or their sequence (video compression).
Compression of 4D LF images is still a relatively unexplored area of research.
Since 4D LF are sequences of 2D images (views), the 2D compression methods may be used to code the views independently.
However, such methods fail to exploit pixel correlations in all four dimensions.
Similar reasoning can be used for 3D methods.
In our first experiment, we were interested in examining the effects of LF compression in three and four dimensions.
To evaluate the compression performance fairly, identical compression method must be used for the 2D, 3D, and 4D case.
Thus, we have created a custom implementation of the JPEG compression method with the ability to process either the 2D, 3D, or 4D data.
Additionally, we are aware of the existence of the JPEG 2000 standard, with the ability to compress the 2D and 3D data in the same manner.
Unfortunately, the JPEG 2000 does not deal with the 4D images.
Since the similarity of adjacent pixels in the third and four dimensions strongly depends on the camera baseline, different results can be expected depending on the baseline distance.
The result of this experiment is shown in Figure~\ref{fig:experiment1}.
In each panel, the horizontal axis shows the bitrate (bits per pixel), whereas the vertical axis shows the mean of the PSNR for multiple rendered focal plane views.

On light fields with a small baseline (Black Fence and Palais du Luxembourg), both 3D compression methods clearly outperform their 2D counterparts over a whole range of bitrates.
Similarly, the 4D JPEG method clearly outperforms its 3D counterpart.
This is not so surprising because pixels at the same spatial position in adjacent views are strongly correlated. 
However, the situation changes with increasing baseline.
With increasing baseline (Lego Bulldozer and Chessboard), adjacent views are less and less similar, which results in higher amplitudes of the underlying transform coefficients.
Consequently, the tide is turning in favor of the less-dimensional compression methods.
Considering the JPEG method, the Lego Bulldozer is a special case because it contains large areas of blackness (black pixels).
It turns out that it is more efficient to compress these solid areas at once using a single 4D block than using multiple 3D blocks.
Similarly, it is more efficient to use a single 3D block than multiple 2D blocks.

\vfill

\begin{figure*}
	\begin{subfigure}[b]{.5\linewidth}
		\includegraphics[width=\linewidth]{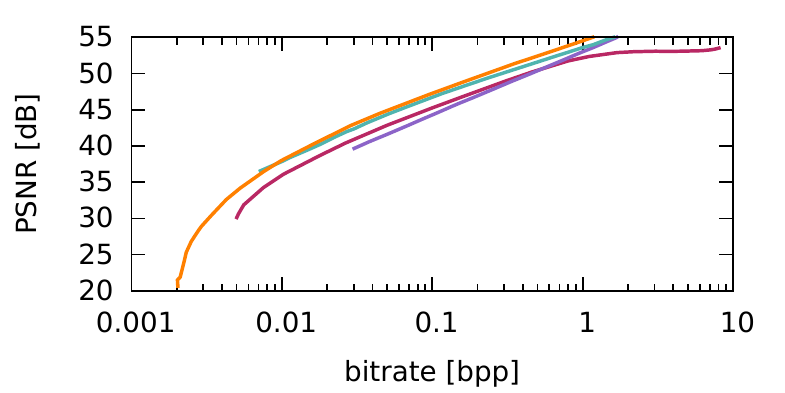}%
		\caption{Black Fence}
	\end{subfigure}%
	\begin{subfigure}[b]{.5\linewidth}
		\includegraphics[width=\linewidth]{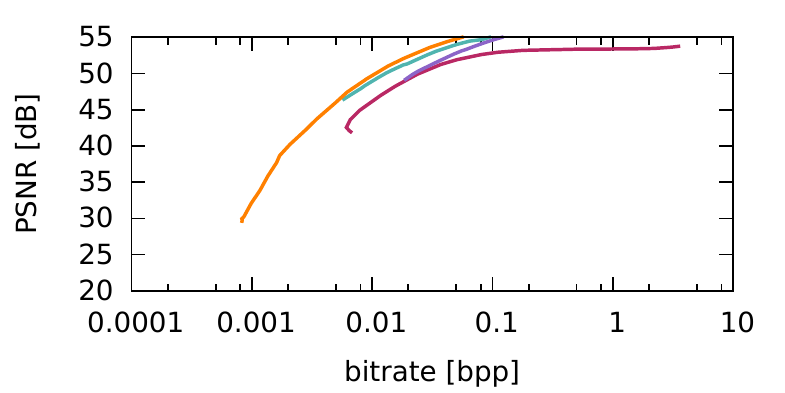}%
		\caption{Chessboard}
	\end{subfigure}%
	\\[15pt]%
	\centering
	\includegraphics[width=.5\linewidth,trim={0 0 0 3cm},clip]{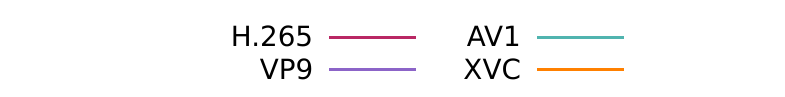}%
	\medskip%
	\caption{
		Experiment 2.
		Performance of video compression methods.
		The XVC and AV1 clearly overcome the older standards.
	}
	\label{fig:experiment2}
\end{figure*}

\paragraph{Experiment 2}
The second thing to notice in the previous section is the employment of the video compression standards.
Upon this, a question arises: whether it be better to compress the 4D light fields as a sequence of 2D frames, or as multi-dimensional body.
We, therefore, measured the performance of all the above-mentioned video compression standards.
The results can be seen in Figure~\ref{fig:experiment2}.
This time results only for two light fields are shown for brevity.
We have, however, got similar results for the other two.
Interestingly, the XVC codec has really shown better compression performance than HEVC and AV1, as claimed by the official website.

\begin{figure*}
	\begin{subfigure}[b]{.5\linewidth}
		\includegraphics[width=\linewidth]{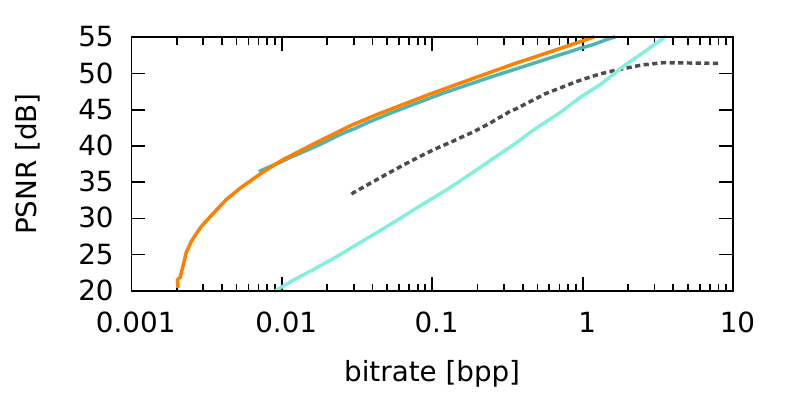}%
		\caption{Black Fence}
	\end{subfigure}%
	\begin{subfigure}[b]{.5\linewidth}
		\includegraphics[width=\linewidth]{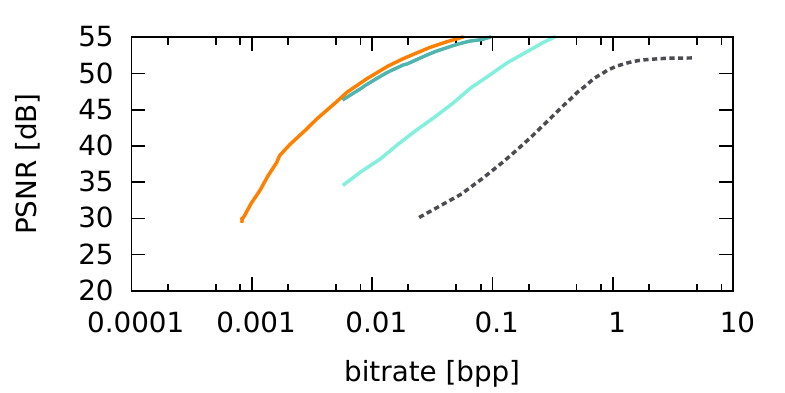}%
		\caption{Chessboard}
	\end{subfigure}%
	\\[15pt]%
	\centering
	\includegraphics[width=.5\linewidth,trim={0 0 0 3cm},clip]{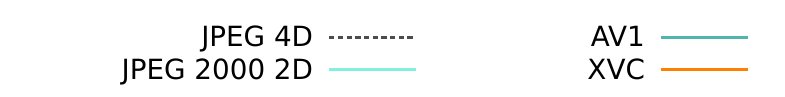}%
	\medskip%
	\caption{
		Overall performance of the best compression methods.
		Video compression methods perform better than all image compression methods.
	}
	\label{fig:overall}
\end{figure*}

To answer the question, "What is the best compression method for LF data?",
we have further compared these results with the best-performing methods from Experiment 1.
The overall comparison is shown in Figure~\ref{fig:overall}.
Interestingly, video compression methods perform better than all image compression methods, even better than their 3D and 4D extensions.

\vfill

\section{Conclusions}
\label{sec:Conclusions}

The purpose of our work was to evaluate the current methods suitable for lossy compression of 4D light fields.
Since the light field is basically a collection of images (views), image compression methods are often the first choice, when it comes to the need for compression.
It turns out that the methods handling the 4D light fields directly in four (or three) dimensions are able to achieve better compression results than common image compression algorithms.
This is, however, dependent on a baseline between neighboring views.
For large baselines (e.g., camera arrays), the common image compression methods come handy.

We have also evaluated the performance of video compression methods.
The underrated XVC compression format demonstrated superior performance, closely followed by the AV1.
This confirms that the latest video compression standards offer better performance than their predecessors.
Eventually, it turns out that these video compression methods perform better than the image compression methods (including their 3D and 4D extensions).

\paragraph{Acknowledgements}
This work has been supported by
the Technology Agency of the Czech Republic (TA CR) Competence Centres project V3C -- Visual Computing Competence Center (no. TE01020415),
the Ministry of Education, Youth and Sports of the Czech Republic from the National Programme of Sustainability (NPU II) project IT4Innovations excellence in science (LQ1602),
and the European Union's Horizon 2020 Research and Innovation Programme under Grant Agreement No 780470.

\bibliographystyle{myabbrvnat}
\bibliography{IEEEfull,sources}

\end{document}